\documentclass[aps,pra,twocolumn,superscriptaddress, a4paper,longbibliography]{revtex4-2}
\usepackage[utf8]{inputenc}

\usepackage{graphicx}
\usepackage{hyperref}
\hypersetup{pdfencoding=auto,colorlinks=true,linkcolor=blue,citecolor=blue}
\usepackage{amsmath}
\usepackage{amssymb}
\usepackage{subfigure}
\usepackage[english]{babel}
\usepackage[normalem]{ulem}

\usepackage{lineno}

\begin{document}

\title{Reusability Report: Comparing gradient descent and monte carlo tree search optimization of quantum annealing schedules}
\author{Matteo M. Wauters}
\affiliation{Niels Bohr Institute, University of Copenhagen, Copenhagen 2100, Denmark}
\author{Evert van Nieuwenburg}
\affiliation{Niels Bohr Institute, University of Copenhagen, Copenhagen 2100, Denmark}
\affiliation{Lorentz Institute and Leiden Institute of Advanced Computer Science, Leiden University, P.O. Box 9506, 2300 RA Leiden, The Netherlands}
\date{\today}

\maketitle

ARISING FROM Yu-Qin Chen et al. Nature Machine Intelligence \url{https://doi.org/10.1038/s42256-022-00446-y} (2022).
\vspace{0.5cm}

The AI system AlphaZero~\cite{Silver_Nature2017,Silver_Science2018} famously learned to play complex and strategic games such as Go and Chess and achieved super-human performance in these tasks. AlphaZero uses a combination of searching through the possible states of a game (Monte Carlo Tree Search) and a neural network to guide that search. The result is an algorithm that learns to choose sequences of moves that lead to a victory. In Ref.~\cite{chen_optimizing_2022}, authors \emph{Chen} et al., set out to employ this same set of techniques to solve a specific class of combinatorial optimisation problems, the 3-satisfiability (3-SAT) problems.


3-SAT is an NP-hard problem, where the task is to decide whether there exists a choice of $n$ binary variables $b_{i=1\ldots n}$ that simultaneously satisfy a set of $m$ clauses of $3$ variables each. 
A clause for a 3-SAT problem is of the form $C = (b_2 \;\textrm{OR}\; \textrm{not}\, b_4 \;\textrm{OR}\; b_6)$. 
The authors focus on the case of $m/n = 3$, for which there are always three times as many clauses as there are variables.
This ratio is smaller than the critical value $m/n\approx 4.2$, above which the ratio of satisfiable expressions drops to zero~\cite{Kirkpatrick_Science1994}, however, it corresponds to a set of hard instances characterised by a unique solution.
The authors provide a dataset that has one or several of such instances for different sizes of 3-SAT.

A 3-SAT problem of $n$ variables can be encoded into a Hamiltonian of $n$ qubits, spanning a Hilbert space of dimension $N=2^n$, whose groundstate provides the solution. 
Solving the 3-SAT problem is then equivalent to finding that groundstate, for which several algorithms exist. 
The algorithm of choice here is the technique of quantum annealing (QA)~\cite{das2005quantum,morita2008mathematical}.

In quantum annealing, finding the groundstate is done by starting in a known (and easily prepared) groundstate of an initial Hamiltonian $H_0$, and then slowly (adiabatically) interpolating to the desired final Hamiltonian $H_f$. 
That is, we perform
\[ H(t) = (1-s(t)) H_0 + s(t) H_f, \]
for $t$ going from $0$ to $T$, and $s(t)$ satisfying $s(0) = 0$ and $s(T) = 1$.
Finding the optimal annealing path -- the actual form for $s(t)$ -- is the central task. 
Starting from the groundstate of $H_0$, and changing $s(t)$ adiabatically will keep the system in the instantaneous groundstate of the full $H(t)$. 
At $t=T$ then, we will have obtained the groundstate of $H_f$ and hence the solution to our 3-SAT problem.
At the same time, going slowly means the annealing takes more time, and we thus have an inherent trade-off between speed and accuracy: the perfect place for an optimisation algorithm to help out.
Following the original paper~\cite{chen_optimizing_2022}, we choose the fidelity as the figure of merit for the accuracy of the algorithm, which simply measures the overlap of the solution obtained with the true solution (which for benchmarking purposes is known). 
A fidelity of $1$ means that the algorithm achieved the perfect solution, whereas a fidelity of $0$ is completely off.
For more general purposes, when the ground state is not known and it might be degenerate or quasi-degenerate, the energy is a better figure of merit. For the specific problem at hand, however, we checked that the fidelity and the energy provide equivalent estimates of the algorithm's accuracy.
\begin{figure}[t!]
    \begin{center}
    \includegraphics[width=8.5cm]{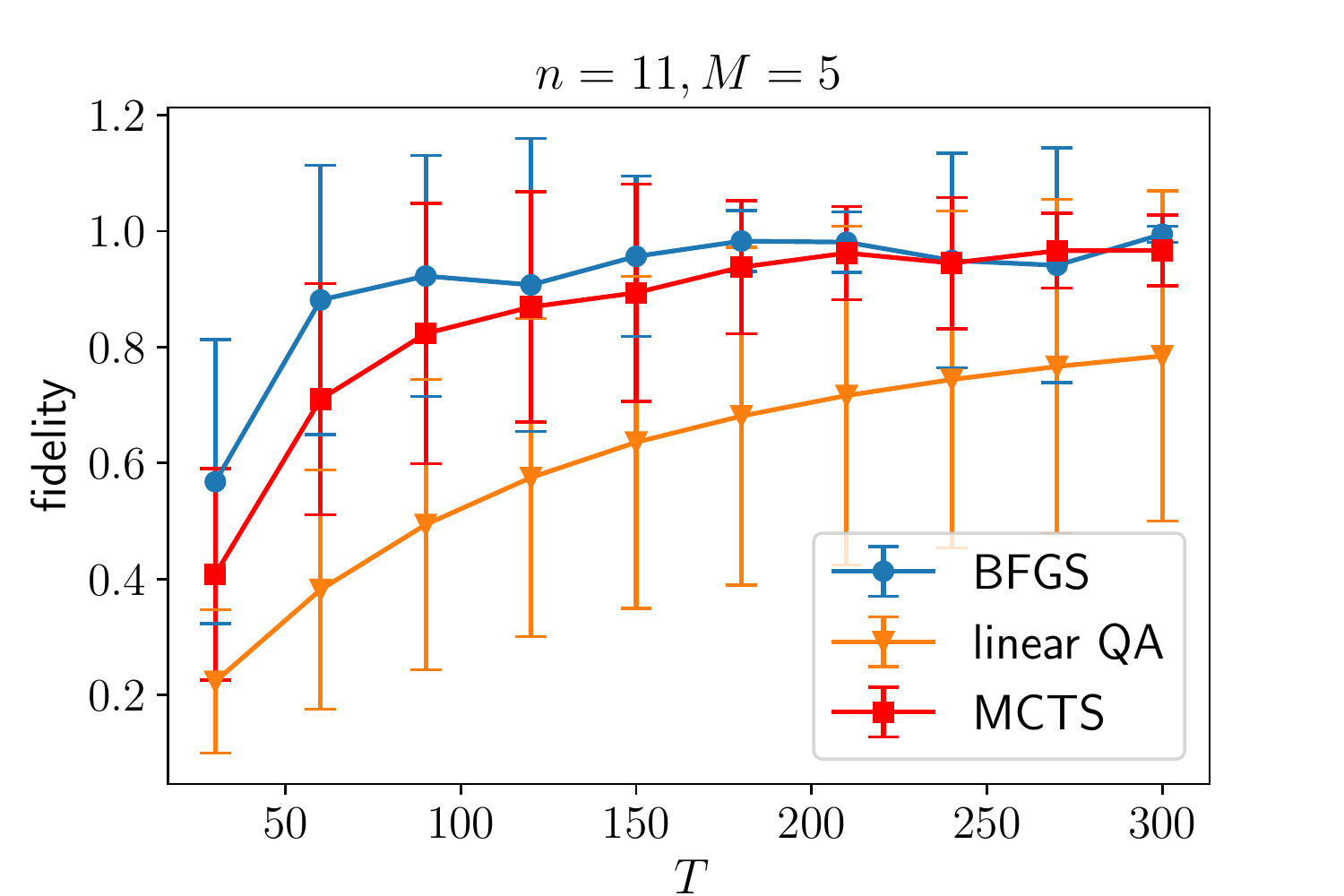}
    \caption{Fidelity at the end of the quantum annealing process vs annealing time $T$ for 3-SAT instances with $n=11$ variables and $M=5$ frequency component in the annealing schedule $s(t)$. We compare a simple linear schedule (orange triangles), gradient descent with the BFGS algorithm (blue circles), and MCTS (red squares). Data are averaged over the 18 instances provided in the GitHub repository; the errorbars are due to the large difference in performance between individual instances.}
    \label{fig:N11_compare}
    \end{center}
\end{figure}

The way \emph{Chen} et al. approach the problem of optimising $s(t)$, is by expanding it as a Fourier series with $M$ frequencies $\omega_k = \pi k /T$, and then optimise the choice of these $M$ Fourier coefficients:
\begin{align}
 s(t) = t/T + \sum_{k=1}^M x_k \sin(\omega_k t).
\label{eq:fourier-components}
\end{align}
The combination of the fixed linear term $t/T$ and the sinusoidal functions ensures that $s(0)=0$ and $s(1)=1$.
The real-valued coefficients $x_k$ were not taken to be continuous values, but they are rather discretised into $40$ steps between $-0.2$ and $+0.2$. 
Finding the right value for each $x_k$ is then a search problem similar to a board game, where the game consists of just $M$ moves, and each move means picking one of the 40 possible values.
This formulation sets the problem up for a solution with an algorithm analogous to AlphaZero.
We keep this structure only for the MCTS optimisation, while for gradient-descent methods we consider the $x_k$ to be continuous and unbounded.

\section{Comparing annealing methods}

In Figure~\ref{fig:N11_compare} we compare different methods for the optimisation of $s(t)$: a linear schedule for $s(t)$, a gradient-based optimiser, and MCTS.
The error bars are obtained as the variance of the optimisation over all 18 provided instances of $n=11$. 
Because the original codebase~\cite{original_code_repository} does not allow for easy reusability of the full MCTS + neural network algorithm, we leave it out of the comparisons for the rest of this work.

\begin{figure*}
    \begin{center}
    \includegraphics[width=\textwidth]{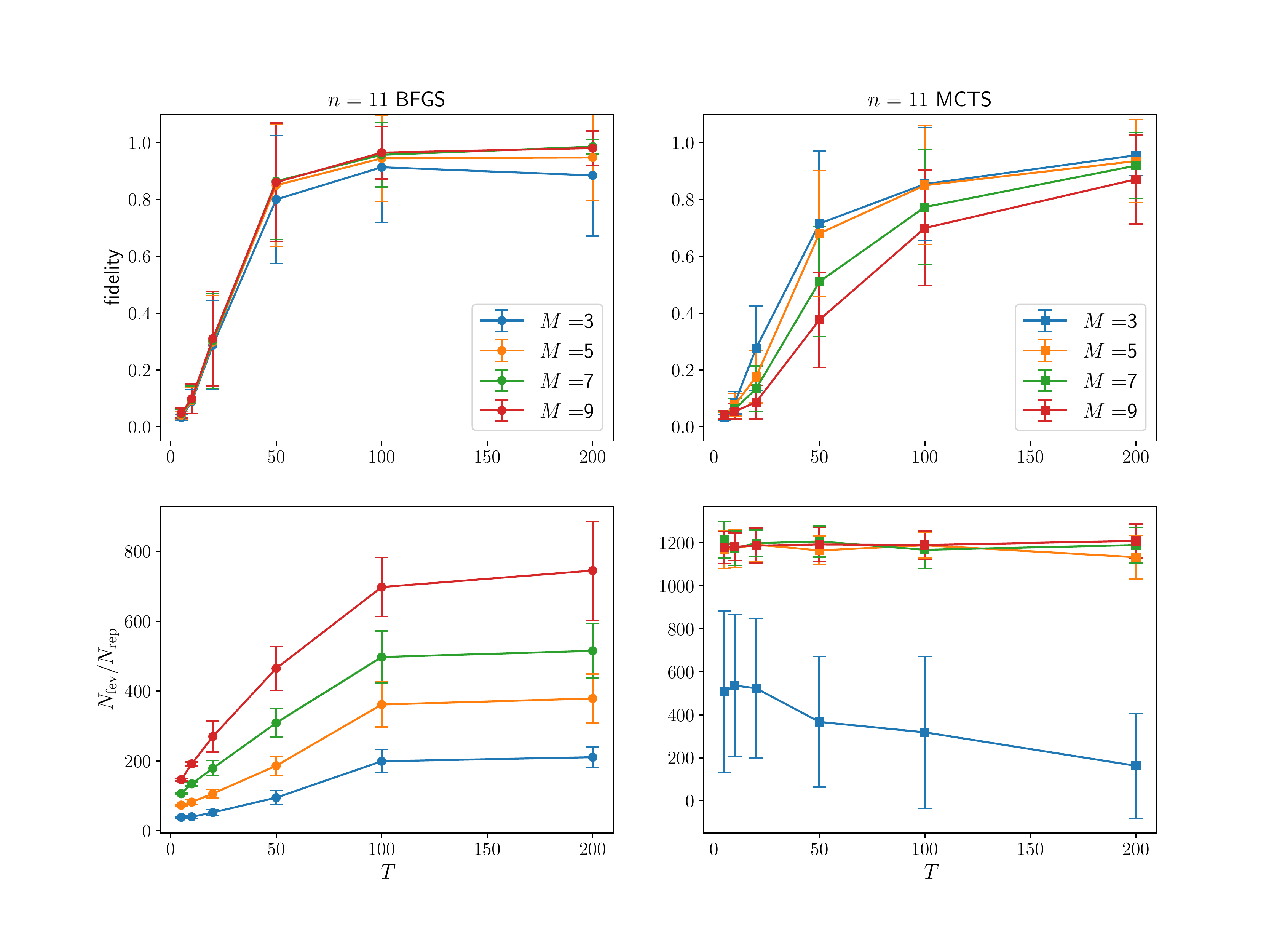}
    \caption{(a)-(b): Fidelity vs annealing time $T$ for different numbers of frequency component $M$ of the annealing schedule. BFGS data are obtained by taking the best result out of 10 local optimisations starting from a noisy linear schedule. 
    Notice that the BFGS fidelity is almost independent of $M$ while for MCTS optimisation the performance slightly decreases for increasing values of $M$.
    (c)-(d) Number of function evaluations required for the schedule optimisation as a function of the annealing time $T$ for different numbers of frequency components $M$. BFGS and MCTS have markedly different behaviours: the former requires an increasing number of queries to the quantum annealer as the annealing time and the cardinality of the parameter space $M$ increases; MCTS instead has a resource requirement that seems independent from $T$ and $M$, beside the data for $M=3$ that reaches convergence much faster.}
    \label{fig:3SAT_fidelity}
    \end{center}
\end{figure*}
Compared to the original figure (Fig. 3a in~\cite{chen_optimizing_2022}), we replace the authors' gradient-based optimisation routine by a more standard implementation. For that, we choose the \texttt{scipy.optimize} version of the Broyden–Fletcher–Goldfarb–Shanno (BFGS) algorithm~\cite{Nocedal_book2006}.
Additionally, for the sake of reusability, we make several more changes to the codebase provided by the authors~\cite{original_code_repository}. 
In particular:
\begin{itemize}
    \item We turn the hard-coded number of frequency components $M=5$ into a parameter.
    \item The convergence criterion for MCTS was set to a fidelity of $0.7$, which we change to either $0.99$ or to whenever the optimisation does not produce a change in the fidelity of more than 1\% over 20 steps.
\end{itemize}
For the rest of this report, we will not include the neural network addition to the MCTS.
In its current form, the codebase does not easily allow such re-use of the full QuantumZero algorithm.

The gradient-based algorithm shown in Fig.~\ref{fig:N11_compare} performs significantly better than the original stochastic descent (SD). 
The author's version of SD evaluates gradients and update $x_k$ sequentially, moving in orthogonal directions in the loss-landscape. 
BFGS evaluates gradients for all $x_k$ and then moves in the direction that minimises the overall cost function.
Even though on average, the BFGS algorithm leads to better fidelities than MCTS, we observed that is more prone to remain trapped in low-quality local minima when the annealing time is large. On individual 3-SAT instances, it can happen that MCTS converges to a better result than BFGS; we observed this for 2 out of 18 instances of the $n=11$ dataset, corresponding to the cases where the linear annealing performs the worst.
Moreover, there is a second merit to using MCTS, which becomes more apparent when we consider metrics such as the number of function evaluations and the achieved fidelity for different numbers of frequency components. 

\section{Frequency component dependence}
We now investigate the dependence of the achieved final fidelity on the number of frequency components used in the expansion of $s(t)$ (see Eq.~\ref{eq:fourier-components}).
We show this result for both the BFGS and MCTS algorithms in Fig.~\ref{fig:3SAT_fidelity}(a)-(b).
The dependence on $M$ for the BFGS optimisation method is only very minimal, whereas for MCTS having more frequency components leads to lower accuracy for a given annealing time $T$. 
Importantly, however, for MCTS the overall fidelity seems to keep increasing for increasing $T$, whereas BFGS tends to get stuck in local minima that prevent it from finding the highest fidelity (c.f. the $M=3$ data for BFGS). 
A better minimum can be found by adding some small noise to the frequency component of the initial guess and repeating the local optimisation to facilitate the exploration of the cost function landscape.
In these results, the MCTS runs ran until convergence was achieved (see above for the code modifications that highlight the convergence criteria). To improve its performance at large $M$, one can choose a more stringent convergence condition, at the likely cost of running the algorithm for a longer time.

\section{Comparing evaluation numbers}
Next, we investigate the required number of evaluations of the annealing schedule. 
This number is a more fair comparison for practical implementations because it dictates how often an annealing experiment would have to be run.
Figure~\ref{fig:3SAT_fidelity}(c)-(d) shows this metric for BFGS and MCTS again for 3-SAT problems with size $n=11$. 
In the former, the number of function evaluations per repetition increases monotonically with both the annealing time and the number of frequency components.
For large $T$ it seems that $N_{\rm fev}$ is reaching a plateau, suggesting that the optimisation process has reached a ``glassy'' phase where the fidelity landscape has a large number of local maxima with small performance difference, with some similarity to the result presented in Ref.~\cite{Buokov_PRX2018}.
The increase of $N_{\rm fev}$ with the number of frequency components is instead due both to the increasing dimensionality of the parameter space and the computational cost of evaluating numerically the gradient of the fidelity.

Interestingly, MCTS shows very little dependence on both $T$ and $M$ of the number of queries to the annealer required for convergence, with the exception of the data for $M=3$. This can be linked to an inherent property of the discretised energy landscape, which might smooth out some of the fine structures present in the continuous space used for BFGS, as well as a better stability of MCTS for large search spaces.
Overall, with the chosen convergence condition, MCTS still require more function evaluations than a single BFGS local minimum search, even though it might reach an advantage over gradient descent for larger system sizes, where large annealing time and schedule optimisation are fundamental for reaching good accuracy.

\section{Max-Cut}
To study the flexibility and universality of the method, we extend the performance analysis on another common classical optimisation problem, namely Max-Cut~\cite{Halperin2004,Zhou_PRX2020} on an unweighted 3-regular graph. Given a regular graph $G=(V,E)$, where $V=\lbrace 1, 2, \dots , N \rbrace$ is the set of vertices and $E= \lbrace \langle i, j \rangle  \rbrace$ is the set of edges, the Max-Cut problem Hamiltonian reads
\begin{equation}\label{eq:Max_cut}
    H_f = \sum_{\langle i,j\rangle \in E} (1 + \sigma^z_i \sigma^z_j ) \ ,
\end{equation}
which corresponds to an antiferromagnetic Ising model on the graph $G$. Since $H_f$ is diagonal in the computational basis, the initial (driving) Hamiltonian can be chosen to be $H_0 = \sum_j \sigma^x_j$ as in the 3-SAT problem investigated in the original paper and in the previous sections of this Reusability Report.

\begin{figure}
    \begin{center}
    \includegraphics[width=8.5cm]{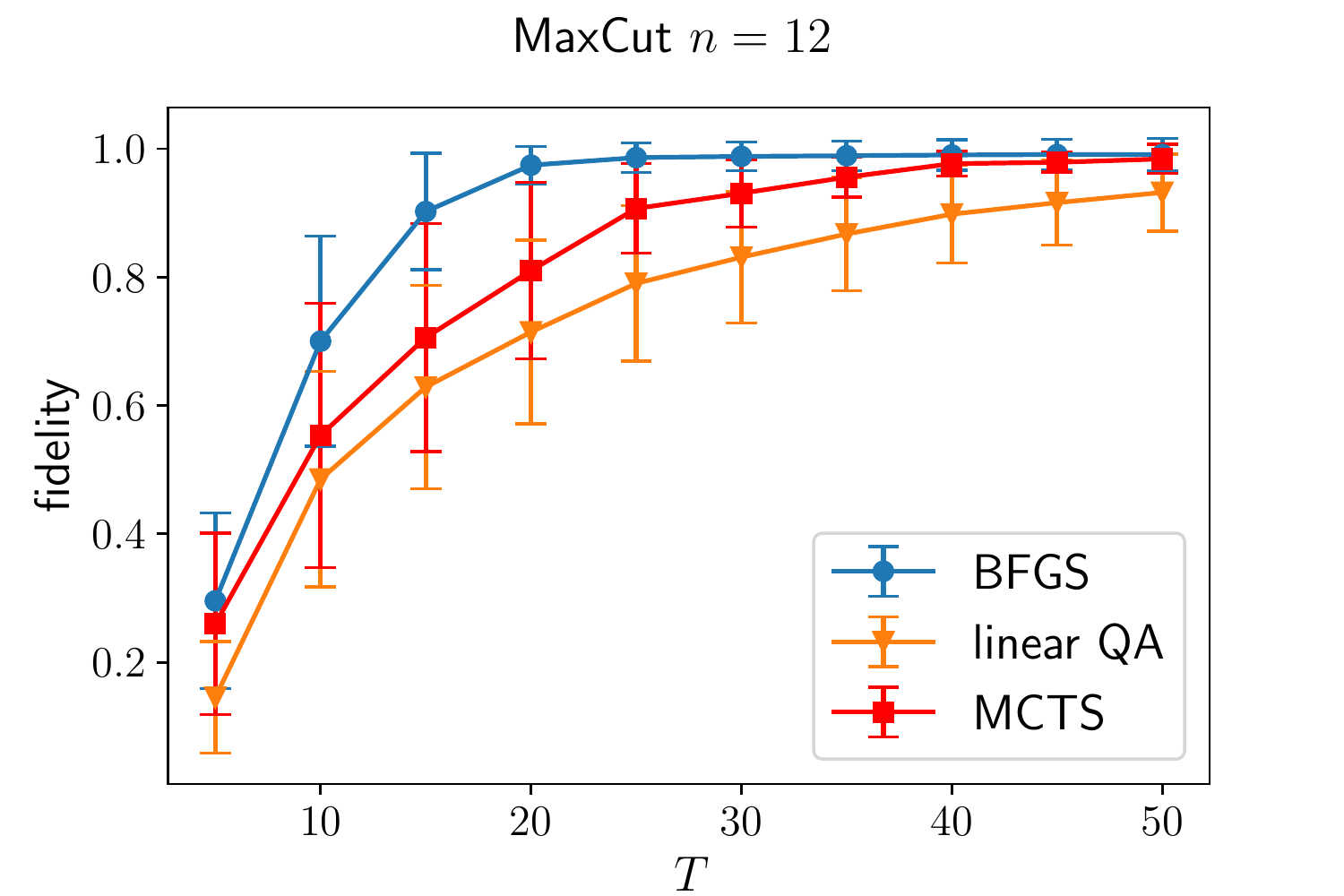}
    \caption{Fidelity at the end of the annealing process vs annealing time $T$ for unweighted MaxCut problems on 3-regular graphs  with $n=12$ vertices. The annealing schedule has been optimised over $M=5$ frequency components. We compare a linear schedule (orange triangles), gradient descent with BFGS algorithm (blue circles), and MCTS (red squares). Data are averaged over the 19 possible connected graphs with the specified geometry.}
    \label{fig:Maxcut_compare}
    \end{center}
\end{figure}

In Fig.~\ref{fig:Maxcut_compare}, we report the fidelity as a function of the annealing time obtained with a single gradient-based optimisation of the schedule $s(t)$ (circles), a single run of MCTS run unitl convergence (squares), and a linear schedule $s(t)=t/T$ (triangles).
The number of vertices in the graph, i.e. the number of qubits, is $n=12$ and each vertex has three neighbors, introducing frustration in the model; the fidelity is averaged over all possible connected graphs with this geometry.
The number of frequency components in the optimised annealing schedule is $M=5$.
We obtain results similar to those presented in Fig.~\ref{fig:N11_compare}: the BFGS optimisation in a continuous variable space leads to better fidelity than MCTS, which, however, shows an improvement over the simple linear schedule.
As observed for the 3-SAT problem, MCTS requires also a larger number of function evaluations to reach convergence: on average, the data reported in Fig.~\ref{fig:Maxcut_compare} required $N_{\rm nfev}\sim 200$ for BFGS and $N_{\rm nfev}\sim 1000$ for MCTS, even though this difference might partially depend on the details of the MCTS algorithm implementation. 
 
\section{Conclusion and Discussion}
In this report, we compared two strategies to optimise the annealing schedule on 3-SAT instances, following the original paper~\cite{chen_optimizing_2022}. 
We considered a ratio between the number of clauses and the number of variables $m/n=3$ and focused on hard instances characterised by a unique solution to the combinatorial problem.
We found that a gradient-based optimisation in a continuous variable space is, in general, preferable over MCTS, both in terms of accuracy and number of calls to the quantum annealer.
This result is in contrast with the original claim and suggests that restricting the optimisation of the frequency components on a discrete set of points limits the accuracy that can be reached for given computational resources. 
However, MCTS displays remarkable independence from both $T$ and $M$ of the number of function evaluations required for convergence. 
Hence, MCTS might have an advantage in large parameter space with complex cost-function landscapes, where gradient-descent optimisation tends to get trapped in local extremes.

\section{Future directions}
Our data suggest that MCTS could become systematically better than gradient descent optimisation when a large annealing time $T$ and a number of frequency components $M$ are needed. 
Typically, this would be the case when the system size $n$ is also large and adiabatic evolution is hindered by vanishing energy gaps.
Hence, a careful scaling analysis of the performance with $N$ can lead to a better understanding of the possible advantages of MCTS as an annealing schedule optimiser. 

Furthermore, an overall improvement could be gained by a different decomposition of the annealing schedule. Our data in Fig.~\ref{fig:3SAT_fidelity} indicates that increasing $M$ leads to very little performance gain; a different basis set might lead to a clearer advantage of MCTS in the large $M$ regime.

Finally, the apparent stability of MCTS when the optimisation problem is ``hard'' (large $T$ and $M$), suggests that it might be a good candidate as the classical optimiser in variational quantum algorithms~\cite{Cerezo_2021}, where gradient-based methods suffer from the appearance of so-called barren plateaus\cite{Uvarov_JPA2021}. 
Recent results~\cite{Mele_2022} showed that this problem can be overcome by transferring smooth optimal schedules. The neural network-aided schedule transfer implemented in the original QuantumZero algorithm might therefore be useful to tackle such issues. 
The current status of the original codebase~\cite{original_code_repository} does not easily allow for such an investigation unfortunately, due to the missing neural network, several hard-coded parameters, and sparsity of comments. 
For that reason we did not cover this in this reusability report.


The modifications and additions we made to the codebase can be found in a separate repository~\cite{code_repository}, which includes a more modular problem setup, more modular annealing methods, the BFGS method, and the code for running the MaxCut optimisation problem.

\section*{Code and Data Repository}
The code (and link to the associated data) can be found at \url{https://github.com/condensedAI/quantumzero} (see ref.~\cite{code_repository}).

\section*{Acknowledgements}
M.W. is supported by the Villum Foundation (Research Grant No. 25310). This project has received funding from the European Union’s Horizon 2020 research and innovation program under the Marie Sklodowska-Curie grant agreement No. 847523 “INTERACTIONS.”

\end{document}